# Mechanisms Inducing Parallel Computation in a Model of *Physarum polycephalum* Transport Networks

Jeff Jones*

*Unconventional Computing Group, University of the West of England, Coldharbour Lane
Bristol, BS16 1QY, UK
email: jeff.jones@uwe.ac.uk*

## ABSTRACT

The giant amoeboid organism true slime mould *Physarum polycephalum* dynamically adapts its body plan in response to changing environmental conditions and its protoplasmic transport network is used to distribute nutrients within the organism. These networks are efficient in terms of network length and network resilience and are parallel approximations of a range of proximity graphs and plane division problems. The complex parallel distributed computation exhibited by this simple organism has since served as an inspiration for intensive research into distributed computing and robotics within the last decade. *P. polycephalum* may be considered as a spatially represented parallel unconventional computing substrate, but how can this 'computer' be programmed? In this paper we examine and catalogue individual low-level mechanisms which may be used to induce network formation and adaptation in a multi-agent model of *P. polycephalum*. These mechanisms include those intrinsic to the model (particle sensor angle, rotation angle, and scaling parameters) and those mediated by the environment (stimulus location, distance, angle, concentration, engulfment and consumption of nutrients, and the presence of simulated light irradiation, repellents and obstacles). The mechanisms induce a concurrent integration of chemoattractant and chemorepellent gradients diffusing within the 2D lattice upon which the agent population resides, stimulating growth, movement, morphological adaptation and network minimisation. Chemoattractant gradients, and their modulation by the engulfment and consumption of nutrients by the model population, represent an efficient outsourcing of spatial computation. The mechanisms may prove useful in understanding the search strategies and adaptation of distributed organisms within their environment, in understanding the minimal requirements for complex adaptive behaviours, and in developing methods of spatially programming parallel unconventional computers and robotic devices.

*Unconventional Computing Group, University of the West of England, UK.

## 1. Introduction

The true slime mould *Physarum polycephalum* is a single-celled organism with a very complex life cycle. The plasmodium stage, where a giant syncytium formed by repeated nuclear division is encompassed within a single membrane, has been shown to exhibit a complex range of biological and computational behaviours.

The plasmodium of *P. polycephalum* is a membrane-bound syncytium of nuclei within a cytoplasm comprised of a complex gel/sol network. The gel phase is composed of a sponge-like matrix of contractile actin and myosin fibres through which the protoplasmic sol flows. Local oscillations in the thickness of the plasmodium spontaneously appear with approximately 2 minutes duration [1]. The spatial and temporal organisation of the oscillations has been shown to be extremely complex [2] and affects the internal movement of sol through the network by local assembly and disassembly of the actin-myosin structures. The protoplasm moves backwards and forwards within the plasmodium in a characteristic manner known as shuttle-streaming.

The plasmodium is able to sense local concentration gradients and the presence of nutrient gradients appears to alter the structure of external membrane areas. The softening of the outer membrane causes a flux of protoplasm towards the general direction of the gradient in response to internal pressure changes caused by the local thickness oscillations. The strong coupling between membrane contraction and streaming movement is caused by the incompressibility of the fluid requiring a constant volume - the weakening of the membrane provides an outlet for the pressure. When the plasmodium has located and engulfed nearby food sources, protoplasmic veins appear within the plasmodium, connecting the food sources. The veins transport protoplasm amongst the distributed extremes of the organism. The relative simplicity of the cell and the distributed nature of its control system make *P. polycephalum* a suitable subject for research into distributed computation substrates. In recent years there have been a large number of studies investigating its computational abilities, prompted by Nakagaki et al. who reported the ability of *P. polycephalum* to solve path planning problems [3]. Subsequent research confirmed and broadened the range of abilities to spatial representations of various graph problems [4, 5, 6], combinatorial optimisation problems [7], construction of logic gates [8] and logical machines [9], [10], and as a means to achieve distributed robotic control [11], robotic manipulation [12] and robotic amoeboid movement [13], [14].

From a pattern formation perspective, *P. polycephalum* can be interpreted as a complex mechanism of dynamical pattern formation based upon the two requirements of efficiency in foraging behaviour (searching of a maximal area) and efficiency in nutrient transport (minimal transport distance and fault tolerance) [15]. The mechanisms used to fulfil these requirements are growth, movement and area reduction. During the growth/foraging stage the plasmodium exhibits a 'default' broadly reticulated outward growth pattern - the homogeneity of the growing plasmodium fragments to form the reticular network [16]. On nutrient rich substrates the growth is typically wave-like and expansive but on nutrient poor substrates the growth is dendritic. There is no strict separation of the two behaviours, however, and both growth types may be observed experimentally in the same environment subject to small local differences in humidity, substrate hardness, substrate roughness or temperature. The foraging plasmodium forms a protoplasmic tube network behind the growth front used to transport nutrients within the organism. Once all nutrients have been located, the topology of the pattern (the protoplasmic tube network) is influenced by the nutrient distribution. The tube network evolves to achieve a compromise between minimal transport costs and fault tolerance [4]. Since the plasmodium obviously cannot have any global knowledge about the initial or optimal topology, the network must evolve by physical forces acting locally on the protoplasmic transport.

In this paper we give an overview of intrinsic mechanisms (model parameters) and external mechanisms (environmental influences) that can influence the behaviour and pattern formation properties of a multi-agent model of *P. polycephalum*. In Section 2 we give an over view of models of *P. polycephalum*. In Section 3 we describe the multi-agent model used in this paper. Intrinsic model parameters affecting network formation are described in Section 4. Environmental mechanisms affecting network formation and adaptation are described in Section 5. In Section 6 we show examples of how such mechanisms can be employed and combined for the approximation of spatially represented computing problems.

## 2. Modelling the Behaviour of *P. polycephalum*

Tero et al. have suggested that protoplasmic flux through the network veins may be cause the physical basis for evolution of the transport network: given flux through two paths, the shorter path will receive more sol flux. By generating an autocatalytic mechanism to reward veins with greater flux (by thickening/widening them) and to apply a cost to veins with less flux (the veins become thinner), shorter veins begin to predominate as the network evolves. This approach was used for the mathematical model of *P. polycephalum* network behaviour to solve path planning problems [17]. This method indirectly supports the reaction-diffusion inspired notions of local activation (strengthening of shorter tubes) and lateral inhibition (weakening of longer tube paths). The starting point for the model of Tero et al. is a randomly connected protoplasmic tube network, surrounding a number of food sources (network nodes) which act as sources and sinks of flux. By beginning with a complete network this method, although successful in generating impressive solutions to network problems, sidesteps the issues and mechanisms of initial network formation and adaptation to a changing nutrient environment.

Gunji et al. introduced a cellular automaton (CA) model which considered both plasmodial growth and amoeboid movement [18, 15]. The model placed importance on the transformation of hardness/softness at the membrane and the internal transport of material from the membrane resulting in movement and network adaptation.

The model was also able to approximate instances of maze path planning and coarse approximations of the Steiner tree problem.

Takamatsu's hexagonal CA [16] mimics the growth patterns displayed under differing nutrient concentrations and substrate hardness. The patterns reflect experimental results well but do not (at least at this stage - oscillatory behaviour is in development) show morphological adaptation as the plasmodium grows. Hickey and Noriega adapted a classical ant colony optimisation algorithm to modify a decision tree in their representation of *P. polycephalum* behaviour in a simple path planning problem [19]. Their algorithm (as with many implementations of ant algorithms) transformed the spatial representation into a graph representation and provided broadly similar results to path optimisation by *P. polycephalum*.

*P. polycephalum* may be interpreted as a spatially represented embodied form of parallel and distributed unconventional computation. In this form of unconventional — or non-classical — computation, the computation is embodied within, and performed by, physical processes within living (or indeed non-living [20]) materials [21]. Unlike the symbolic algorithms used to control classical computing devices, problems and their solutions are represented as spatial patterns. The process of computing the problem is typically performed by propagation of information throughout the substrate until a final or stable state is reached. The distributed control and the simple parallel nature of its component parts comprising *P. polycephalum* render it an attractive candidate material for synthetic collective computation and soft-robotics applications. In order to utilise this 'material' for useful computation we must find mechanisms and methods to induce morphological adaptation. Such mechanisms will lead to the ability to 'program' the material to solve useful problems. AS *P. polycephalum* is a living organism, it suffers from problems of speed, repeatability, and the unpredictability and relative fragility associated with biological systems. A suitable model is required, however one which is also composed of relatively simple parts, utilises local interactions and demonstrates distributed control of its behaviour.

## 3. A Multi-agent Model of *P. polycephalum*

Although slime mould has enviable computational properties, it also has limitations due to the fact that it is a living organism. Although relatively easy and inexpensive to culture, slime mould is also relatively slow (certainly compared to silicon computing substrates) and must be maintained within strict environmental parameters of temperature, light exposure and humidity. Slime mould may also be relatively unpredictable in its behaviour. Although the unpredictability is useful in certain circumstances it can be a hindrance when repeatable measures of its performance are required. We therefore require a synthetic analogue of slime mould. One technique available is computer modelling, where we attempt to reproduce the complex patterning of slime mould along with the complex interactions it has with its environment.

It is important to note, however, that we are not simply trying to extract the features of slime mould for classical algorithms. Such an approach may indeed prove useful for certain tasks, but would not inform us in any way about the distributed emergent behaviour and control of the organism. Instead what we wish to do is to construct a virtual material using the same principles (and apparent limitations) of slime mould. Namely, simple component parts and local interactions. The aim is to generate collective emergent behaviour utilising self-organisation to yield an embodied form of material computation which can reproduce the wide range of complex patterning and environmental responses seen in slime mould.

In [22] we introduced a large population of simple components, mobile agents, (a schematic view of a single agent is shown in Fig. 1 whose individual behaviour was coupled to the other agents via a diffusive chemoattractant lattice. A further extension to the model was presented in [23] which enabled adaptive population sizes, where the population size adapted automatically to the local availability of nutrients in the environment. Each agent corresponds to a small fragment of plasmodium gel/sol structure. Agents respond to the concentration of a hypothetical 'chemical' in the lattice, orient themselves towards the locally strongest source and deposit the same chemical upon making a single step forwards. Although individually the particle behaviour is very simple, the collective behaviour is emergent and complex, exhibiting self-organised pattern formation. The population represents both the structure (population global pattern) and flux (population movement) within the *P. polycephalum* plasmodium.

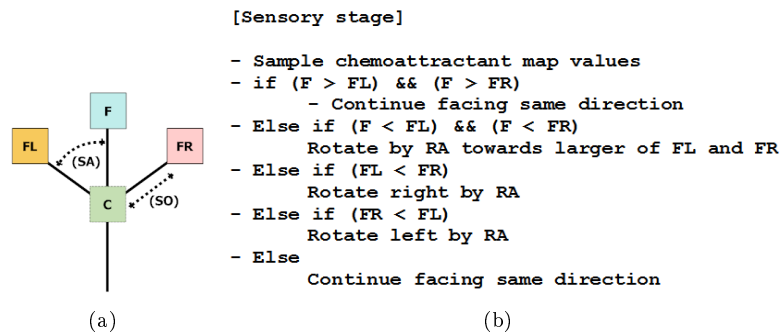

Fig. 1. Base agent particle morphology and sensory stage algorithm. (a) Illustration of single agent, showing location 'C', offset sensors 'FL','F','FR', Sensor Angle '*SA*' and Sensor Offset '*SO*', (b) simplified sensory algorithm.

The collective movement trails left by agent movement spontaneously formed emergent transport networks which underwent complex evolution, exhibiting emergent minimisation and cohesion effects under a range of sensory parameter settings (Fig. 2). To simulate the presentation of nutrients and repellents to the agent population discrete stimuli were projected into the chemoattractant lattice. Positive values represented attractant stimuli (i.e. nutrients) and negative values (below zero) represented repellent stimuli. The effect of nutrient stimuli is to attract nearby agents and the network adaptation is constrained by the attraction to the nutrients. The resulting network structures (nutrient stimuli representing graph vertices and network trails representing edges) show minimisation behaviour and formed proximity graphs with the same properties as those observed in *P. polycephalum* [23] (as an example see the stabilised networks in Fig. 6). Repellent stimuli cause the networks formed by the agent population to avoid repellents, with the resultant networks approximating plane division problems such as Voronoi diagrams [24]. The collective behaves as a virtual material demonstrating characteristic network evolution motifs and minimisation phenomena seen in soap film evolution, for example the formation of Plateau angles, T1 and T2 relaxation processes and adherence to von Neumann's law (Fig. 3).

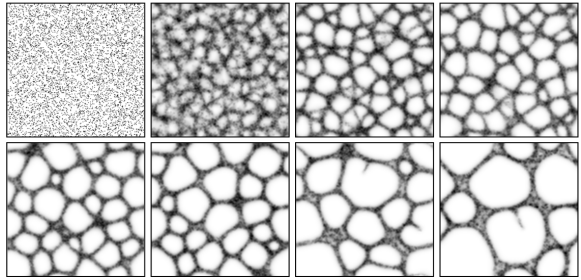

Fig. 2. Spontaneous formation and evolution of transport networks. Lattice 200×200, 6000 agent particles, *SA* 22.5°, *RA* 45°, *SO* 9.

## 4. Model Parametric Mechanisms Inducing Changes in Morphological Adaptation

Different parametric settings within the model (although *not* changing the underlying algorithm for each agent) can affect the pattern formation properties of the model, altering both network structure and the evolution of the networks, even without the effect of external environmental stimuli.

### 4.1. *Variations in Sensory Parameters*

Variations of the agent sensor angle and agent rotation angle (See schematic of a single agent particle in Fig. 1) result in a wide range of reaction-diffusion type patterns [25] (see Fig. 4 for an overall visualisation of the SA/RA parameter space). How might these different patterns, all arising from the same particle behaviour, relate to *P. polycephalum*? One possible relationship is the different pattern types seen in *P. polycephalum* under different environmental conditions, such as substrate hardness and nutrient concentration which both affect the patterning properties of the organism [16].

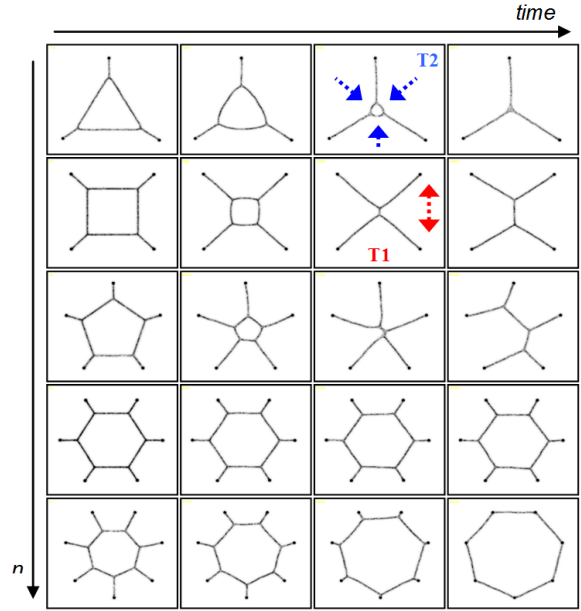

Fig. 3. Material adaptation of the multi-agent model demonstrates $T2$ and $T1$ relaxation processes (arrowed), Plateau junction angles, and adherence to von Neumann's law. $n$ represents number of nodes and edges.

At low SA/RA settings (e.g. SA 22.5°, RA 45°) the networks continuously form branching paths from existing edges and reconfigure their structure over time (Fig. 5,a-d). Higher values (e.g. SA 60°, RA 60°) result in minimising networks where the number of cycles and edges are reduced over time (Fig. 5e-h). At higher values (e.g. SA 90°, RA 45°) result in networks disassembling to form discrete island patterns (Fig. 5i-l).

### 4.2. *Variations in the Scaling Parameter*

The effect of the sensor scale parameter on graph evolution is also significant. Larger sensor offset distance (SO, in pixels) results in thicker network paths, faster network evolution, and coarse-grained networks, where each edge encompasses many data points (Fig. 6a). Decreasing the SO parameter results in more fine-grained graphs with more edges and cycles connecting nearby data points (Fig. 6b and c).

### 5. Environmental Mechanisms Inducing Changes in Morphological Adaptation

Graph transformation in classical computing is typically achieved by changing the algorithm operating on a set of source data (for example data points representing graph vertices). Different algorithms operating on the same dataset will construct different graphs, depending on the particular algorithm, resulting in a final static output graph. If the algorithm changes, or if the set of data points changes, the algorithm must run again to generate a new output graph. Computation by *P. polycephalum*, however, is a seamless and dynamical response to changing environmental conditions. These changes are communicated to the organism by means of physical stimuli. Although these stimuli can take many forms (for example direct physical stimuli, thermal, optical, gravitational) the most studied are chemoattractant and chemorepellent stimuli. For this reason the most influential cues to stimulate network adaptation in the model are the diffusing chemoattractant and chemorepellent gradients in the model lattice.

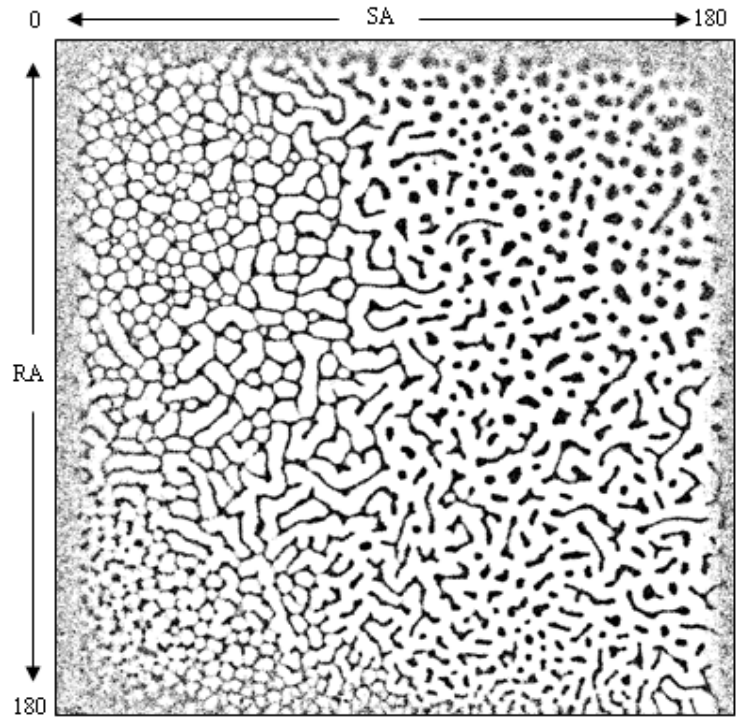

Fig. 4. Parametric mapping of *SA* and *RA* sensory parameters yields a wide range of Turing-type reaction-diffusion patterns.

## 5.1. *Addition and Removal of Nutrients*

The model plasmodium, like the real organism, runs continually and changes to the spatial configuration of nutrient data points are propagated throughout the environment by simulated mechanisms of diffusion. Addition of nutrients adds a new source of attractant to the lattice. Because the propagation relies upon diffusion, the speed of network adaptation is limited by the propagation speed.

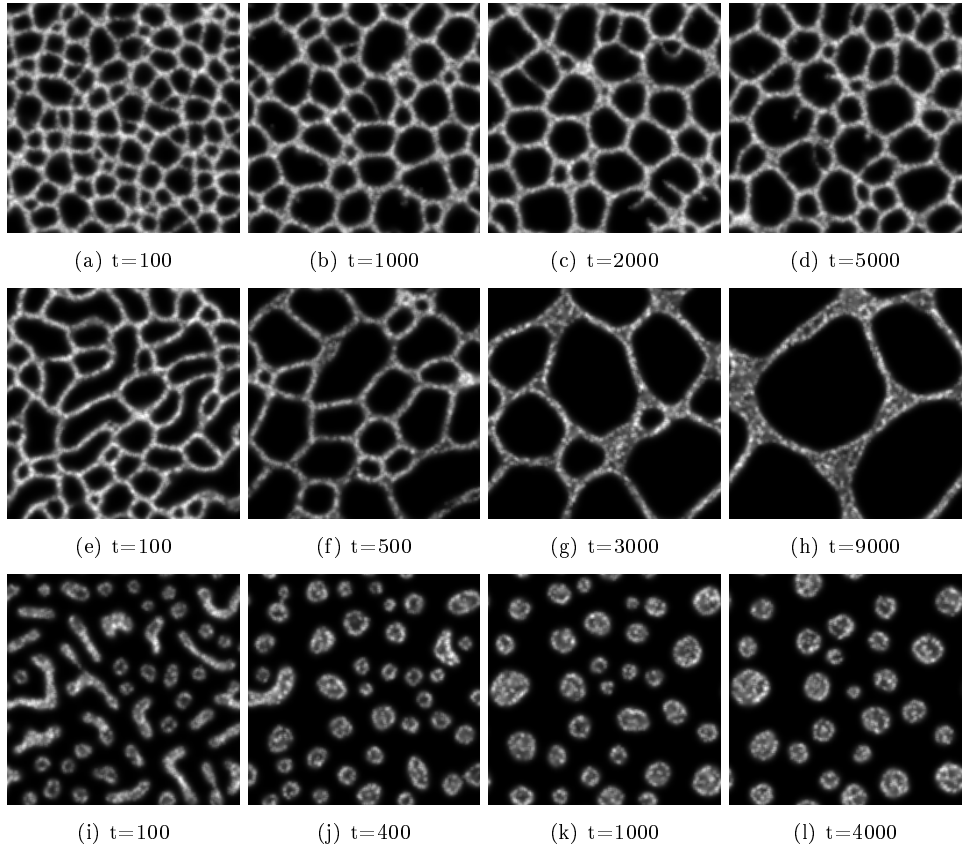

(a) t=100     (b) t=1000     (c) t=2000     (d) t=5000

(e) t=100     (f) t=500     (g) t=3000     (h) t=9000

(i) t=100     (j) t=400     (k) t=1000     (l) t=4000

Fig. 5. Effect of agent sensor angle (SA) and rotation angle (RA) on network type. 6000 particles inoculated on $200 \times 200$ lattice with no environmental stimuli and using periodic boundary conditions. Network evolution proceeds from left to right, a-d) SA $22.5°$, RA $45°$, e-h) SA $60°$, RA $60°$, i-l) SA $90°$, RA $45°$.

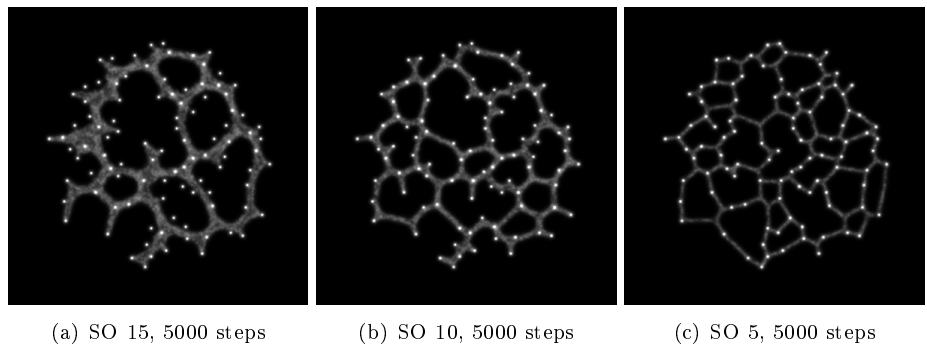

(a) SO 15, 5000 steps     (b) SO 10, 5000 steps     (c) SO 5, 5000 steps

Fig. 6. Effect of sensor scale on network structure. particles inoculated at random positions amongst 100 nodes and final pattern recorded after 5000 scheduler steps. a) SO 15 gives very coarse-grained networks which pass between paths of nearby nodes, b) SO 10 gives thicker paths and fewer cycles, c) SO 5 results in fine-grained networks passing through the data points.

Growth of *P. polycephalum* is dependent on the availability, concentration and placement of nutrient sources in the environment. The plasmodium membrane is sensitive to diffusion gradients and preferentially grows towards nearby sources of nutrients by extending pseudopodia towards the nutrients. On nutrient poor conditions (damp filter paper and oat flake food nodes) *P. polycephalum* initially constructs a spanning tree when inoculated at a single site and constructs networks relating to the upper ranges of the Toussaint hierarchy of proximity graphs when inoculated at multiple sites [6]. To assess the effect of chemoattractant diffusion, nutrient concentration and nutrient placement on the growth in the particle model we inoculated a small population at a simulated food source - a stimulus value projected onto the diffusion field at regular intervals. We assumed (as in [6]) that any food sources covered by the model plasmodium would suppress the diffusion of chemoattractant from that source. Fig. 7 shows the effects of placing food near the plasmodium. When the diffusing chemoattractant gradient reaches the initial site of population initialisation, the particles closest to the gradient are attracted towards the gradient and move towards it. The movement of the population stimulates division at the periphery of the collective and a pseudopod-like process emerges and moves towards the source of food. The width of the pseudopodium active region is dependent on the size of the chemoattractant gradient. When the food source is reached the engulfment by the particles suppresses the diffusion from the node and the connection is stabilised and minimised. The adaptation also occurs when stimuli are removed from the environment by retracting the pseudopodium from the deleted source and adapting the network shape in response (Fig. 7, bottom row, deleted node is circled).

### 5.2. *Avoidance of Light Irradiation and Repellents*

Avoidance of light irradiation and chemorepellents was implemented in the model by decreasing the chemoattractant detected by the agent sensors by multiplying by a weighting factor in areas of the arena exposed to values corresponding to light exposure ($L_d$, weight factor from 0 to 1, default of 0.2) and repellents ($R_d$, weight factor from 0 to $-1$, default -0.2) respectively. Fig. 8 illustrates the response of both fixed and adaptive population sizes to simulated light hazards, in this case vertically placed bars obstructing a straight path between two nutrient sources. For fixed populations the population shifts the bulk of its shape away from illuminated regions (Fig. 8, left). In adaptive populations the model plasmodium curves around the light obstacles to connect the two nutrient sources at opposite sides of the arena (Fig. 8, right), thus reproducing the collision-free path planning by *P. polycephalum* reported in [26].

When growing in an environment containing both nutrients and repellents the model avoided the repellent regions (light squares) whilst growing towards and consuming nutrients (dark squares) in the arena (Fig. 9).

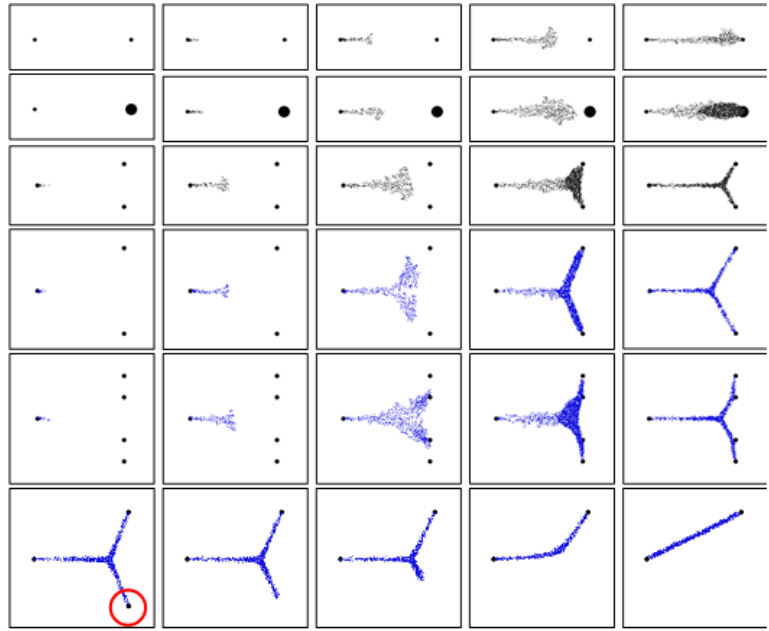

Fig. 7. Discovery, pseudopod extension, tube network formation, pseudopod retraction and network adaptation in particle model. Population initialised on left-most nutrient source, evolution proceeds from left to right. Food source on right projects chemoattractant into the diffusing gradient field. Pseudopod extension observed as particles multiply. Network minimisation continues once all nodes have been located.

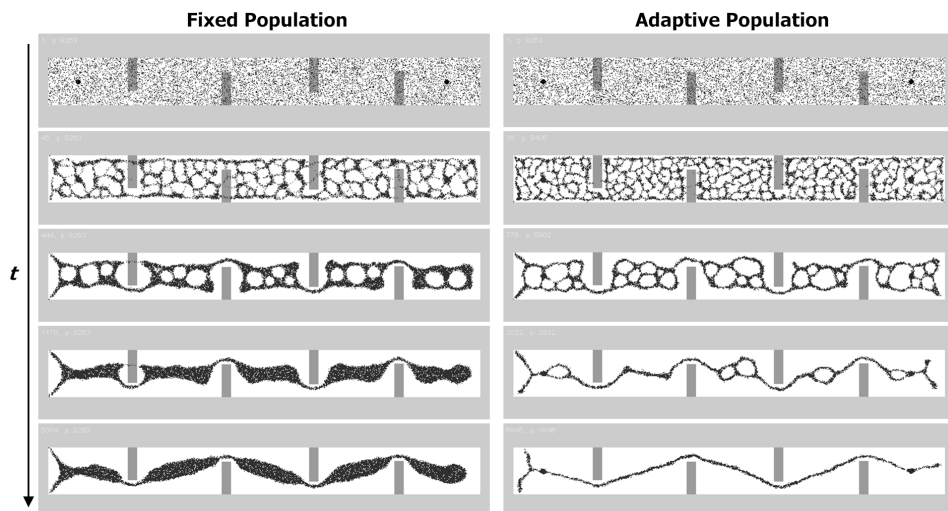

Fig. 8. Avoidance of simulated light hazards in fixed and adaptive populations. Population (particle positions shown) is initialised in arena with two nutrient sources at either end interrupted by vertical bars of projected light. Population adapts to avoid migrating onto exposed areas. (left) fixed population, (right) adaptive population.

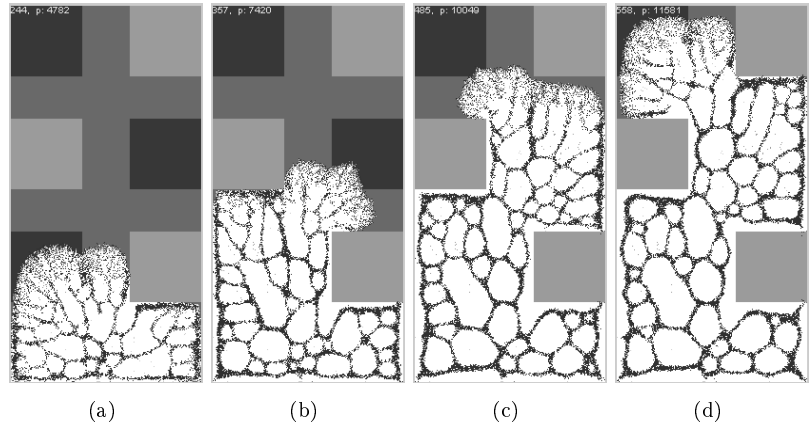

Fig. 9. Growth towards attractants and avoidance of repellents in the model plasmodium. (a-d) Growth of model plasmodium (particle positions shown) on simulated nutrient substrate with high concentration regions indicated by dark grey squares and repellent regions by light grey squares.

### 5.3. *The Presence of Passive Obstacles*

In the previous example the removal of a nutrient source occurred in open space as changes in the diffusive lattice were propagated towards the particle network. However the retraction of the virtual pseudopodia also occurs when the environment is patterned with obstacles which cannot be occupied (in *P. polycephalum* such areas include agar covered in dry plastic film, such as the 'dead-ends' in the mazed in [3]). Even in the presence of obstacles, however, the pseudopodia can retract from areas devoid of nutrients as demonstrated in Fig. 10 in which the model slime mould is initialised in a T-shape which contains only two nutrient sources at the extremal points of two channels. At the other channel the collective forms a pseudopodium which retracts away from the region devoid of stimuli, following the channel bounded by the obstacles (Fig. 10,c-d). Eventually this pseudopodium completely retracts and merges with the stable path connecting the two nutrients (Fig. 10,e).

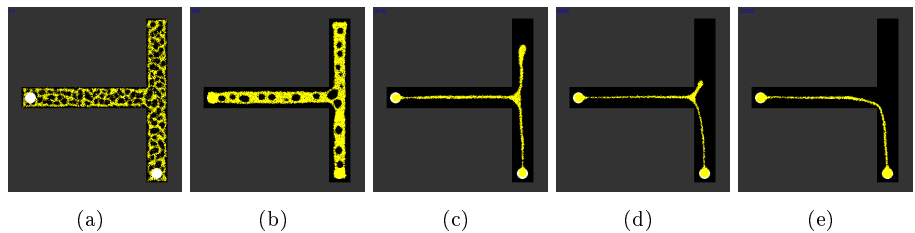

Fig. 10. Pseudopodium withdrawal constrained by passive obstacles. (a) model plasmodium is inoculated in a T shape with two nutrient sources (white discs) and uninhabitable background (grey) and the initial transport network forms, (b-e) Network adaptation removes smaller channels and withdraws the pseudopodium from nutrient free areas, leaving only the nutrient sources connected.

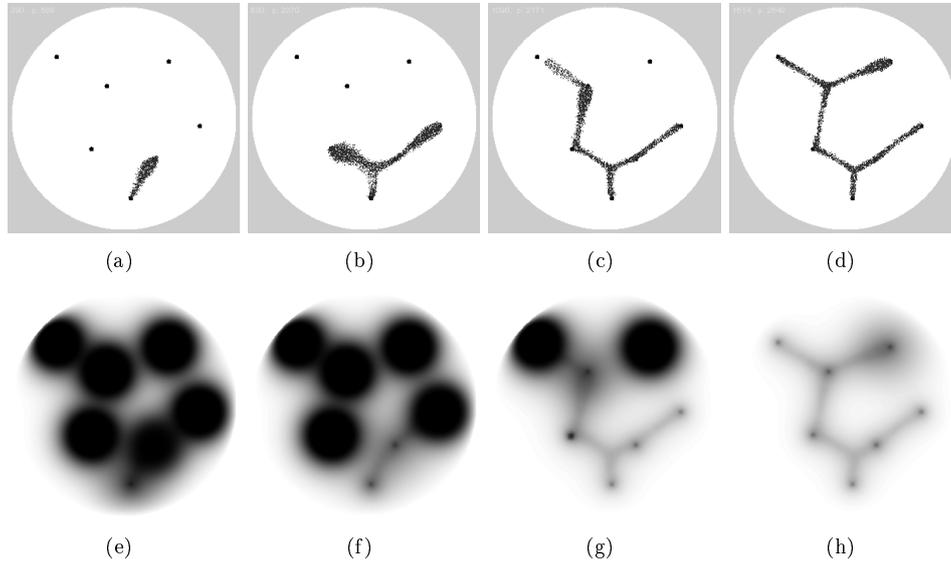

Fig. 11. Construction of a spanning tree by model plasmodium. (a) Small population (particle positions shown) inoculated on lowest node (bottom) growing towards first node and engulfing it, reducing chemoattractant projection, (b-d) Model population grows to nearest sources of chemoattractant completing construction of the spanning tree, (e-h) Visualisation of the changing chemoattractant gradient as the population engulfs and suppresses nutrient diffusion.

### 5.4. *Dynamical Effects Caused by the Engulfment of Nutrients*

The importance of environmental stimuli must emphasised in the virtual material approach. Without any external stimuli the virtual material simply reproduces dynamical reaction-diffusion patterning. It is the stimuli provided by external attractants or hazards which force the material to adapt its spatial behaviour. The environmental stimuli are used to specify problem configuration and the final pattern of the material in relation to the stimuli represents the problem solution. The specific mechanism utilised is the diffusion of attractants (or repellents) within the environment. The presence of these stimuli at the periphery of the material provides the impetus for its morphological adaptation. The interaction between environment and the material is two-way, however. When the material migrates towards and engulfs a nutrient source, the diffusion of nutrients from that source is suppressed. This changes the local configuration of chemoattractant gradients (as demonstrated by the changing concentration gradient profiles in Fig. 11e-h as a spanning tree is constructed using the virtual material) which ultimately changes the spatial pattern of stimuli offered to the material. This mechanism is an efficient use of the environment as a spatial storage medium and 'offloads' some complex computation to the environment. This may explain the reason why slime mould, and its virtual material representation, can perform such complex behaviours without requiring complex nervous system or indeed any neural tissue.

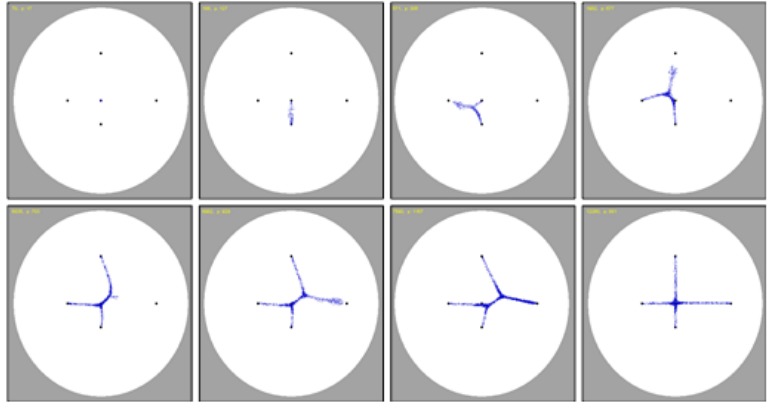

Fig. 12. Adaptive tube network formation as model plasmodium senses diffusing chemoattractant gradients at increasing distances. Inoculation of particle population at central node. Closest food source is 50 pixels away. Distances to other nodes are 75, 100 and 125 pixels. Transport network shape adapts as new food sources are discovered. Final image shows optimised transport network when foraging is complete.

## 5.5. *Nutrient Distance*

As the growing population discovers new food sources the transport network automatically adapts its shape in response to the new nutrient sources. Fig. 12 shows the network adaptation as information about the food locations (in the form of diffusing chemoattractant gradient fields) arrives at the inoculation point at different times due to their increasing distance. The network adapts in response to the changing environment by extending pseudopodia towards the nutrient sources as they are encountered and constructing a transport network connecting the nutrients. The collective then minimises the network distance when all the food is located.

## 5.6. *Nutrient Angle*

The attraction of the multi-agent transport networks to discrete stimuli in the lattice causes the network adaptation to be constrained as the networks are 'snagged' at the stimulus locations. The constraining effect is dependent on the angle between the outer two nutrient points connected to the central nutrient point (Fig. 13). At large angles, the two flows connecting the outer points pass through the central point (Fig. 13, top). As the angle is decreased, however, the two separate particle flows become closer to each other (Fig. 13, middle). Eventually the two flows merge and, if the attraction of the flows is greater than the attraction to the central point, the network detaches partially, in a zipping motif, from the central point [23]. A new path connects the central point to the zipping paths and a Steiner point is formed at this new junction (Fig. 13, bottom). The position of this point eventually stabilises at a location which minimises the distance between all three nodes. Note that if the Steiner point is still relatively close to the original node, a large increase in this node concentration can 'unzip' the network, removing the Steiner point and re-attaching the two outer nodes directly to the central node.

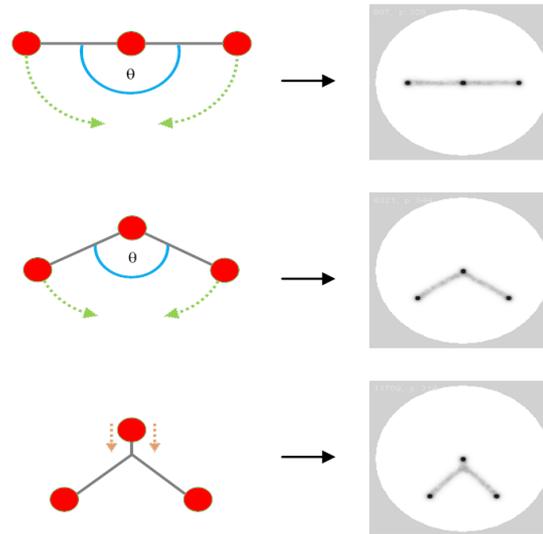

Fig. 13. A critical angle between nutrients affects network adaptation. Top: Three nutrient nodes in a line at 180° results in no minimisation, Middle: decreasing the angle $\theta$ brings the two flows connecting either side of the node closer together, Bottom: a critical angle exists where the two flows merge and the network detaches partially from the node, causing the formation of a Steiner point (a junction where the three paths connect).

The critical angle is dependent on both the Sensor Offset (SO) parameter and the concentration of the nutrient stimuli. Larger SO parameter values result in thicker network paths, with the two paths meeting and merging more quickly than narrower paths. High concentration stimuli causes stronger attraction of the network paths to the nutrient sources, whereas low concentration stimuli causes less adherence to the original nutrient locations. Network adaptation caused by this zipping phenomena is not isolated to single triads of nodes: The detachment by zipping from a node (and thus changes in neighbouring node path angles) can cause subsequent zipping of nearby nodes, provoking minimisation of the entire network structure. The minimisation of the entire network stabilises when the effects of von-Neumann's law (the number of surrounding nodes in a cycle, see Fig. 3) prevents further network minimisation.

### 5.7. *Nutrient Size and Concentration*

The effect of nutrient size and nutrient concentration was studied in the particle model by inoculating a small population at the centre of a circular arena surrounded by four food sources at identical distances (Fig. 14). Each image in the figure shows the concentration gradients (left side of each image) and the particle positions (right side). When nutrient sources were the same size and concentration the particle collective grew by extending pseudopodia towards the food sources at the same time (Fig. 14, left column). When the pseudopodia reached the nutrients the engulfment suppressed the projection of nutrients into the arena and reduced the concentration gradients. When the nutrients were at different concentrations the pseudopodia were preferentially extended in the direction of the strongest nutrients first. Pseudopodia were only extended to the remaining nutrients (in decreasing order of concentration) when the gradient of the previous nutrient block was suppressed by engulfment. When all sources had been located network adaptation took place (Fig. 14, middle column). When the nutrients varied in size only, the concentration gradients from the larger nutrient blocks were larger and pseudopodia were again extended to the nutrients with higher concentrations first. After all nutrients had been discovered the network adapted again to cover the regions of the nutrients but the mass of the population was shifted towards the position of the largest nutrient block (Fig. 14, right column, bottom image).

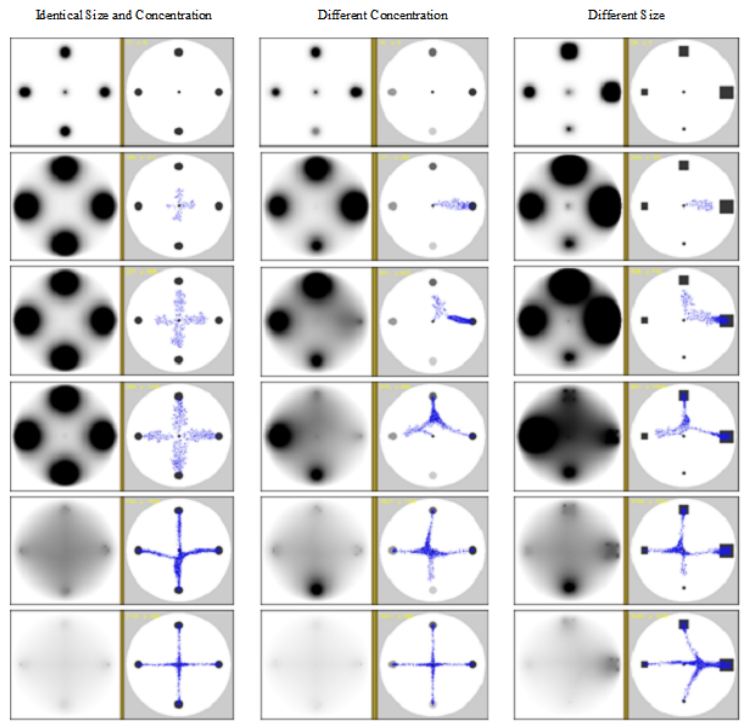

Fig. 14. Foraging behaviours affected by nutrient concentration and size. Evolution proceeds from top to bottom. Left of each image shows diffusion gradients, right side shows food and particle positions, Left: All food is of equal size and concentration, Middle: Food is of equal size but different concentration, Right: Food is equal concentration but different size.

### 5.8. *Nutrient Concentration and Consumption*

The suppression of nutrient concentration gradients represents a complex non-linear and dynamical environment as the nutrient gradients are in constant flux. In the real world the complexity is further compounded by the consumption of nutrients. Even the quality of the nutrients can affect the growth of the plasmodium patterns [27]. The effects of such complex interactions between spatial position, concentration and consumption in the model are shown in Fig. 15. The examples show an initial inoculation site at the bottom centre of a circular arena. Above the inoculation site are three nutrient sources of identical size but potentially different concentration (pixel intensity 255 or 50). In Fig. 15a) all three nutrients are identical and the pseudopodium grows towards the closest source and extends further pseudopodia as the middle nutrient is consumed and gradients from the side nutrients reach the collective. A similar situation occurs in Fig. 15b) but the migration from the central nutrient is delayed because the nutrient is of higher concentration than the outer nodes and its consumption takes longer. In Fig. 15c) both outer nutrients are of higher concentration and, although the collective initially grows towards the closer central node, the mass of the collective extends towards the outer nodes as the gradient is stronger. In example Fig. 15d) the growth of the collective towards the central nutrient node is waylaid by the stronger attraction towards the left node. Extension towards the rightmost node only occurs after the majority of the left node has been consumed. Finally, in Fig. 15e) the collective grows towards the (closest) central node and then to the leftmost node which is equal in strength. Growth towards the node on the right only occurs when the other nodes are mostly depleted.

### 5.9. *Background Substrate Concentration*

The emergent transport networks formed by the microscopic interactions of the particle population with their environment reflect not only static pattern morphologies adopted by the *P. polycephalum* plasmodium, but also the long term network adaptation seen in the organism. *P. polycephalum* morphology, evolution and behaviours are strongly affected by the availability, location and concentration of nutrients. The organism appears to behave in a manner which initially optimises (maximises) area exploration and which later adapts its network by optimising (minimising) network distance and network resiliency to damage. The growth and adaptation morphology of *P. polycephalum* also depends on the nutrient concentration of the growth substrate itself - rich background environments (for example, oatmeal agar) generate florid wave-like radial expansive growth whereas nutrient poor environments (for example, dampened filter paper) result in tree-like dendritic growth. Although attempting to characterise the behaviour of such different growth types runs the risk of anthropomorphism, the wave-like (nutrient rich) behaviours appear more aggressive in terms of the apparent speed of growth and rapid area coverage. The dendritic (nutrient poor) behaviours appear almost tentative in terms of growth

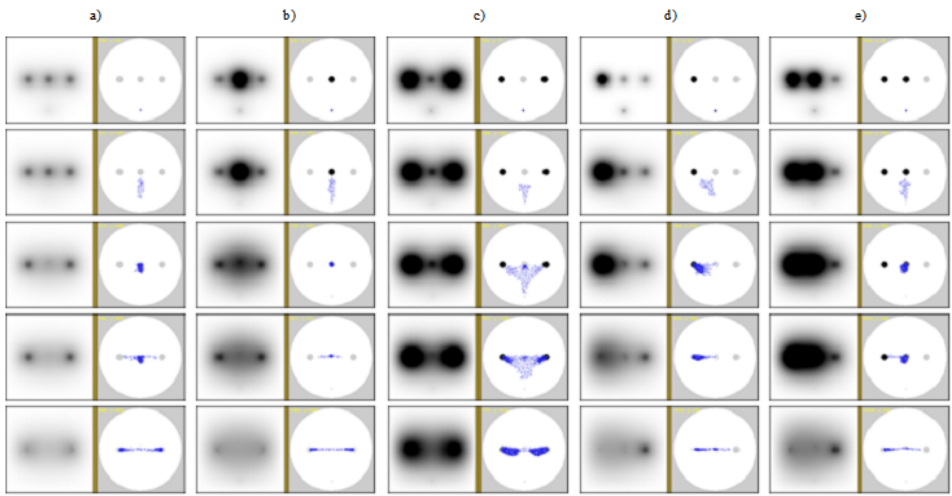

Fig. 15. Effect of nutrient concentration and consumption on foraging behaviours. Evolution proceeds from top to bottom. Left of each image shows diffusion gradients, right side shows food and particle positions.

speed and area coverage.

The same morphological and apparent behavioural effects were observed in the particle model when background environmental conditions were modified. Fig. 16 and Fig. 17 show the effects of high and low concentrations of background nutrient substrates on the morphology and collective behaviours of the particle population. The environment is represented by discrete oat flakes (white circles) and nutrient rich oatmeal agar medium background (grey background). In both experiments identical geometric configurations were used in the environment (i.e. the placement of simulated oat flakes were identical), but the concentrations of the background substrate were different. In both cases the population grows as the environment is searched for nutrients. After the search is complete, and background nutrients exhausted, both conditions spontaneously undergo network contraction and minimisation until minimal network configurations are achieved. In the high concentration condition (Fig. 16) the growth is wave-like and expansive and the final network configuration resembles a relative neighbourhood graph with a number of cycles in the network. The growth in low concentration background condition (Fig. 17) shows dendritic growth patterns and the search of the environment is slower. The final network configuration is also more tree-like with only a small inner cycle.

The cause of the differences in growth and adaptation patterns is the background nutrient concentration. In the high concentration condition, the background presents a stronger stimulus to the periphery of the model plasmodium and the 'pull' of the environment causes expansive movement outwards and provides space for growth. As nutrients are depleted by the outwardly moving population, the background stimulus moves further outward from the edge of the collective and the outer regions of the population move further towards the nutrients. In the low concentration condition the lower background concentration does not provoke such a strong attraction to the population because the nutrient gradients are approximately the same as the background level of chemoattractant flux. Growth of the population does gradually occur outwards but this is only when significant differences in concentration are created by the local consumption of nutrients by foraging particles. The effect of nutrient concentration on the particle population size can be observed in Fig. 18 which indicates the rapid expansion in population size under high concentration conditions (the environmental search is completed at a maximum population size of 21258 at 470 scheduler steps) followed by a rapid initial collapse in population size as network adaptation continues. The low concentration condition shows a slower rise in population size (maximum population of 13574 with search completed in 3440 steps) with a slower initial rate of network adaptation. The population size in both conditions converges within 15,000 steps although the final network size of the low concentration condition is slightly smaller than the high concentration condition due to the reduced number of cyclic regions in the configuration.

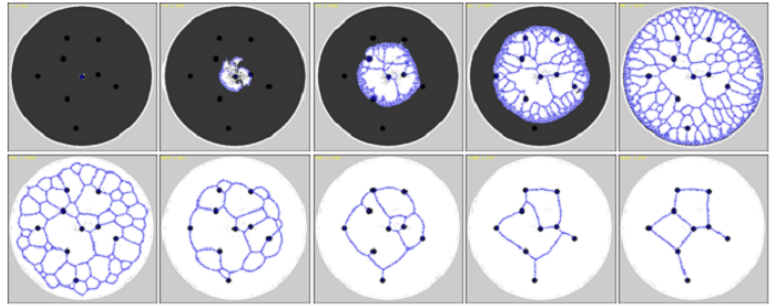

Fig. 16. Wave-like network expansion and cyclic adaptation configuration in high background nutrient concentration. Background nutrient concentration: 0.01, Population initialised at central node. Images sampled at 11, 110, 213, 327, 487, 1476, 4576, 7004, 11240 and 25000 scheduler steps, Consumed nutrient indicated by white areas.

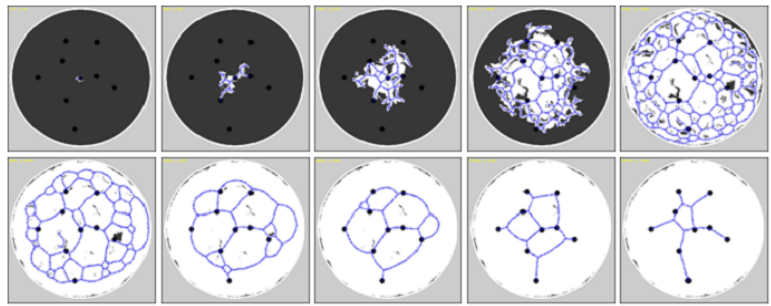

Fig. 17. Tree-like network expansion and adaptation with lower background nutrient concentration. Background nutrient concentration: 0.001, Consumed nutrient indicated by black areas, Population initialised at central node. Images sampled at 116, 566, 1262, 2361, 3636, 4880, 7448, 9640, 17620 and 36708 scheduler steps.

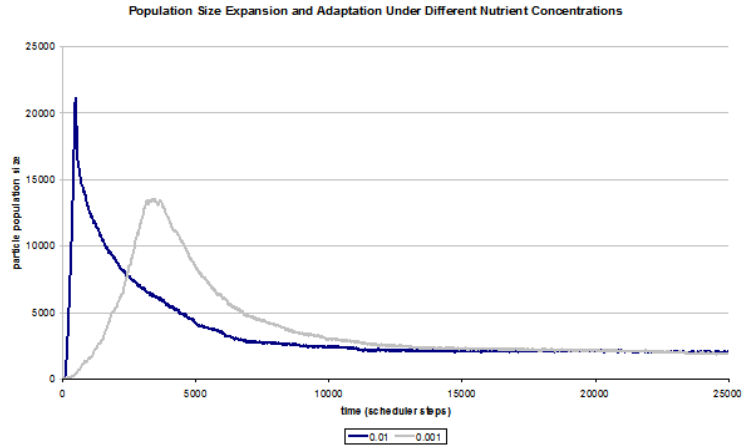

Fig. 18. Plot of population growth and adaptation at different nutrient strengths. Concentration 0.01 indicated by sharp peak in population size, 0.001 indicated by lighter shade and lower peak in population size.

The total number of particles created (during the entire evolution of the experiment) in the high concentration condition (8.7 million) is less than the low concentration condition (9.4 million), reflecting the search efficiency of the radial expansive growth pattern compared with the dendritic foraging observed in the low concentration condition. This can be seen in Fig. 17 where the dendritic search often 'misses' nearby food (unconsumed food shown as lighter shade) whereas the high concentration condition has located all of the food resources during the completion of network expansion.

When inoculated at a single food source and surrounded by isolated nutrients — with no stimuli from the background substrate — the behaviour of the model plasmodium mimics that of the real organism by locating nearby sources of food as the chemoattractant gradient from each source propagates outwards, surging towards the nutrients with pseudopodium growth, engulfing them, and constructing a network which approximates a Steiner tree (Fig. 19, bottom). The foraging and minimisation behaviour of the model closely approximates the behaviour of *P. polycephalum* (as shown in Fig. 19, top).

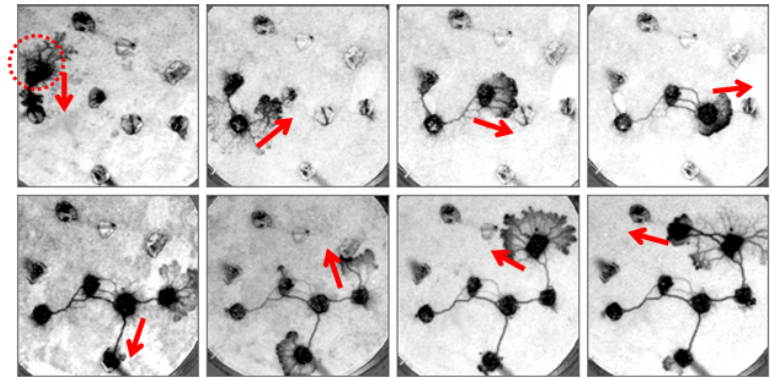

(a)

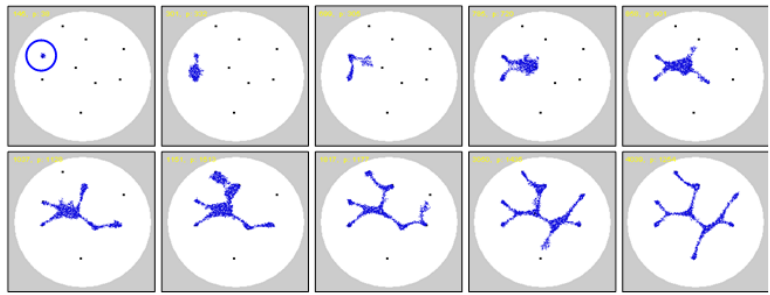

(b)

Fig. 19. Approximation of spanning tree by pseudopod extension in nutrient poor conditions in *P. polycephalum* and model. Top: Plasmodium is initialised on circled oat flake. Pseudopodia extend from original source towards nearby oat flakes and a protoplasmic tube network connects the food sources. Arrows show current active growth front for clarity and the image contrast was enhanced with standard operators to aid viewing of protoplasmic tubes, Bottom: Food nodes indicated as dots inside circular arena approximating shape of Petri dish. Particles shown as mass of dots, model plasmodium inoculated on left side (circled node). Images show foraging and engulfment of nodes projecting diffusing chemoattractant gradients. Final network configuration approximates the Steiner tree.

## 6. From Morphological Induction Mechanisms to Unconventional Computation

How can the innate pattern formation of the model, and the mechanisms which influence it, be put to use for unconventional computational purposes? We must select problems which can be presented in terms of spatial patterns. The solution of the problems can be instigated by using the low-level mechanisms in Sections 4 and 5 to guide the evolution of the particle transport networks towards the desired outcome. The solution is, in-turn, represented by the persistent global state of the network.

The low-level mechanisms have been presented in isolation in each section but they can also be combined (for example combinations of attractants and repellents, or combinations of high and low concentration stimuli) to give the necessary computation. The innate pattern formation and influencing mechanisms may also require different initialisation mechanisms and be subject to external control systems in order to guide the evolution towards the desired problem outcome. In this section we present an overview of example problems that can be approximated by the material adaptation approach using some of the mechanisms described in Sections 4 and 5.

### 6.1. *Proximity Graph Problems*

Approximation of proximity graphs is a natural application of the model, since *P. polycephalum* networks approximate these graphs [28, 6]. The model transport networks also approximate graphs in the Toussaint hierarchy [29], the particular type depending on nutrient concentration but typically match Relative Neighbourhood Graphs [23]. The model may be initialised on a single node, constructing the graphs as it detects and engulfs the nodes (as in Fig. 11 and Fig. 19). Alternatively the model may be initialised at random positions. After initialisation a network self-assembles between the data points and subsequently evolves to minimise the number of edges (Fig. 20a). When initialised as a solid mass of particles and undergoing gradual shrinkage, the 'blob' of model plasmodium minimises down the Toussaint hierarchy to construct a Spanning Tree at high nutrient concentration (Fig. 20h) or a Steiner Tree at low nutrient concentration (Fig. 20i).

### 6.2. *Convex Hull*

The Convex Hull of a set of points is the smallest convex polygon enclosing the set, where all points are on the boundary or interior of the polygon. Classical algorithms to generate Convex Hulls may be inspired by intuitively mechanical methods, such as shrink wrapping an elastic band around the set of points, or rotating calipers around the set of points [30, 31]. It is possible to approximate the convex hull using the model plasmodium by initialising the population as a circular ring of model plasmodium outside the points (Fig. 21), to represent a deformable elastic material. This bounding 'band' then shrinks to encompass the outer region of the set of points. The minimising properties of the paths ensure that the edges of the hull are straight and convex. There are some practical limitations of this approach. Firstly, the bounds of the set of points must be known in advance. Secondly, points which are inside the hull, but close to the 'band' (for example near the top edge in the third image of Fig. 21) may attract the band inwards, forming a concavity.

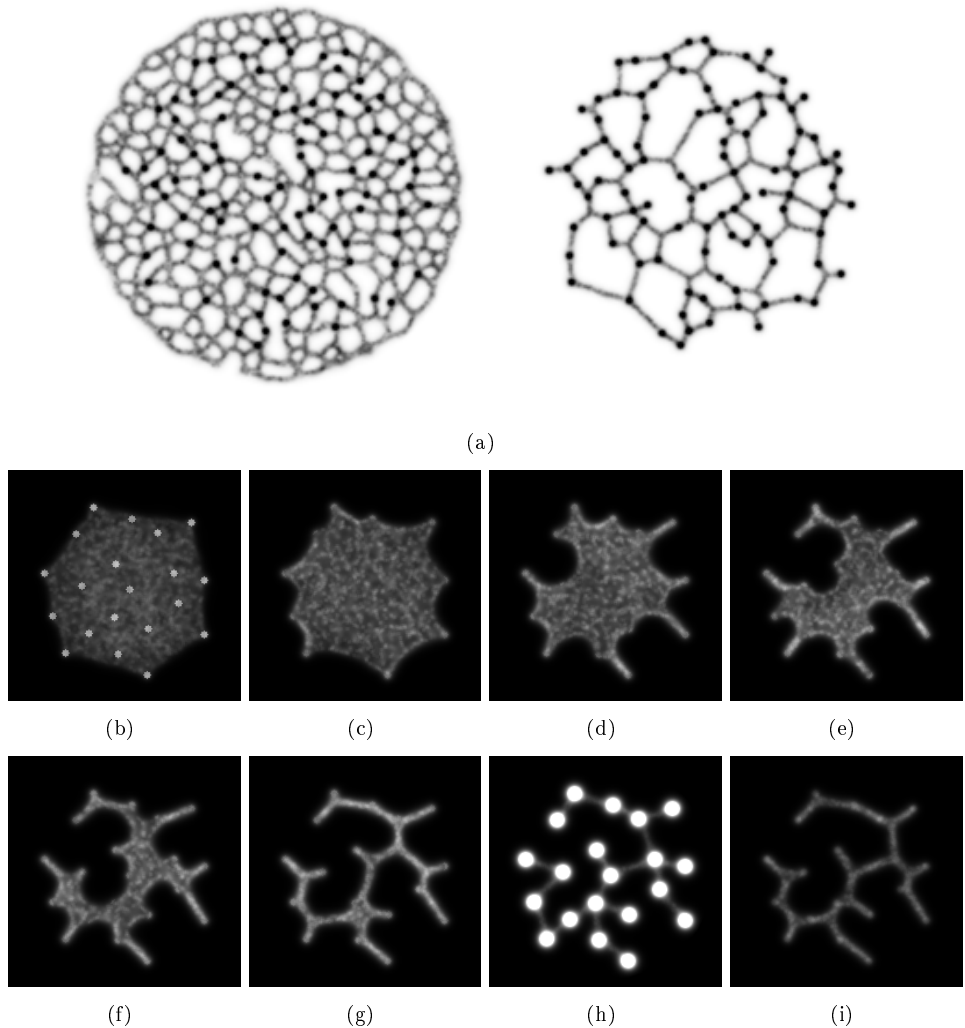

Fig. 20. Proximity graph formation using the multi-agent model. a) initialisation at random positions results in a self-assembled network (left) which minimises to approximate the Relative Neighbourhood Graph (right), b) Initialisation as a solid blob (20 data points indicated by white dots), c-g) slow shrinkage of the model plasmodium results in adaptation of the blob to the data points, h) at high nutrient concentration a spanning tree is formed, i) at low nutrient concentration a Steiner tree is formed.

This may be avoided by restricting the nodes to deposit stimuli only when directly touched by the encircling band.

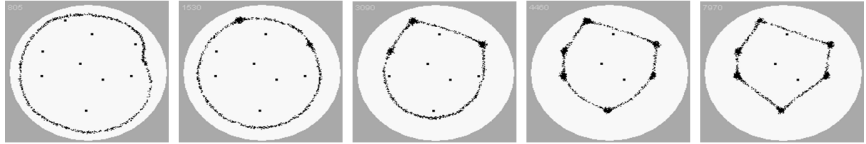

Fig. 21. Approximation of Convex Hull by shrinking band of model plasmodium. A circular band of model plasmodium initialised outside the region of points shrinks. In this example nodes only emanate nutrients when touched by model plasmodium (see text). Contact points of Convex Hull are indicated by larger nodes.

Alternatively it is possible to have the 'band' shrink around the array of points which are actually repulsive to the band. This is achieved by projecting a repellent source at all nutrient node locations. The band will shrink towards the nodes but will not actually touch the nodes. This generates a hull which encompasses the nodes but does not directly touch them (Fig. 22).

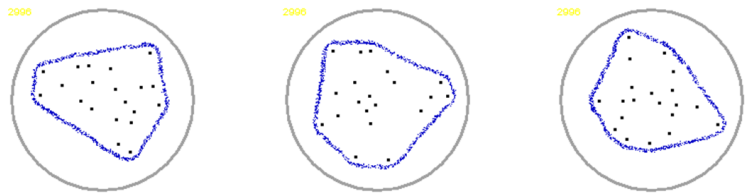

Fig. 22. Convex Hull via shrinkage around hazard stimuli. Three separate examples are shown. A band of model plasmodium shrinks around the set of points to approximate the Convex Hull. Note a small peripheral region is indicated because of the repulsive region.

If the boundary of the hull points is not known then it is possible to utilise a method which employs self-organisation and repulsion to approximate the hull, as shown in Fig. 23. In this approach the particle population is initialised at random locations within the lattice. The particles are repulsed by the chemorepulsive nodes and move away from these regions. If a particle touches a node it is annihilated and randomly initialised to a new blank part of the lattice. Over time, the inner region of the lattice becomes less frequently habited by particles but in contrast the outer region (whose border is away from the repulsive region) becomes more populous. The increasing strength of the emerging hull trail attracts particles from inside the region (because the deposited 'ring' of flux is more attractive than the inner region) and the particles are drawn out into this ring. The natural contraction of the ring approximates the final convex hull.

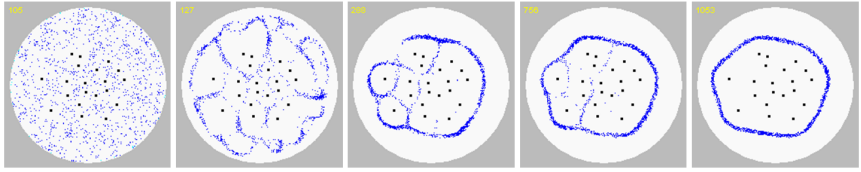

Fig. 23. Convex Hull via self-organisation within repulsive field. Particle population is initialised randomly in the arena and is repulsed by nodes. Convex hull emerges at the border and internal connections gradually weaken.

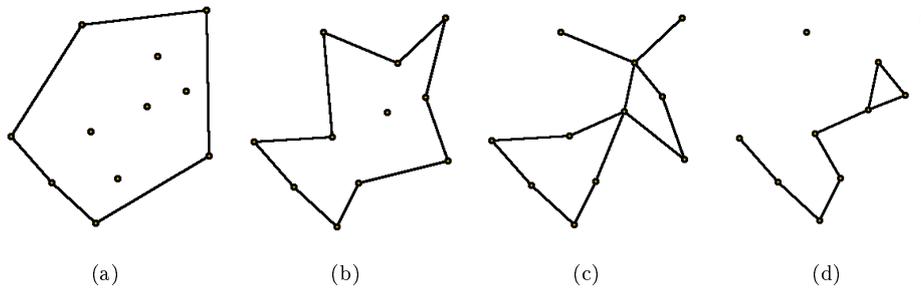

<div align="center">(a)       (b)       (c)       (d)</div>

Fig. 24. Examples of $\alpha$-shape of a set of points as $\alpha$ decreases.

### 6.3. *Concave Hull*

The Concave Hull is the area occupied by, or the 'shape' of, a set of points is not as simple to define as its convex hull. It can be considered as the minimum region (or footprint [32]) occupied by a set of points, which cannot, in some cases, be represented correctly by the convex hull [33]. For example, a set of points arranged to form the capital letter 'C' would not be correctly represented by the convex hull because the gap in the letter would be closed (Fig. 25a). Attempts to formalise concave bounding representations of a set of points were suggested in the definition of $\alpha$-shapes [34]. The $\alpha$-shape of a set of points, $P$, is an intersection of the complement of all closed discs of radius $1/\alpha$ that includes no points of $P$. An $\alpha$-shape is a convex hull when $\alpha \to \infty$ (Fig. 24a). When decreasing $\alpha$, the shapes may shrink, develop holes and become disconnected (Fig. 24b-d), collapsing to $P$ when $\alpha \to 0$. A concave hull is non-convex polygon representing area occupied by $P$. A concave hull is a connected $\alpha$-shape without holes.

The model plasmodium approximates the concave hull via its automatic morphological adaptation as the population size is reduced. The reduction in population size may be implemented by adjusting the growth shrinkage parameters to bias adaptation towards shrinkage whilst maintaining network connectivity. In the examples shown below the model plasmodium is initialised as a large population within the confines of a Convex Hull (calculated using an algorithmic method) of a set of points (Fig. 25b). By slowly reducing the population size, via biasing parameters towards shrinkage, the model plasmodium adapts its shape as it shrinks. Retention to nodes is ensured by chemoattractant projection and as the population continues to reduce, the shape outlined by the population becomes increasingly concave (Fig. 25c-f).

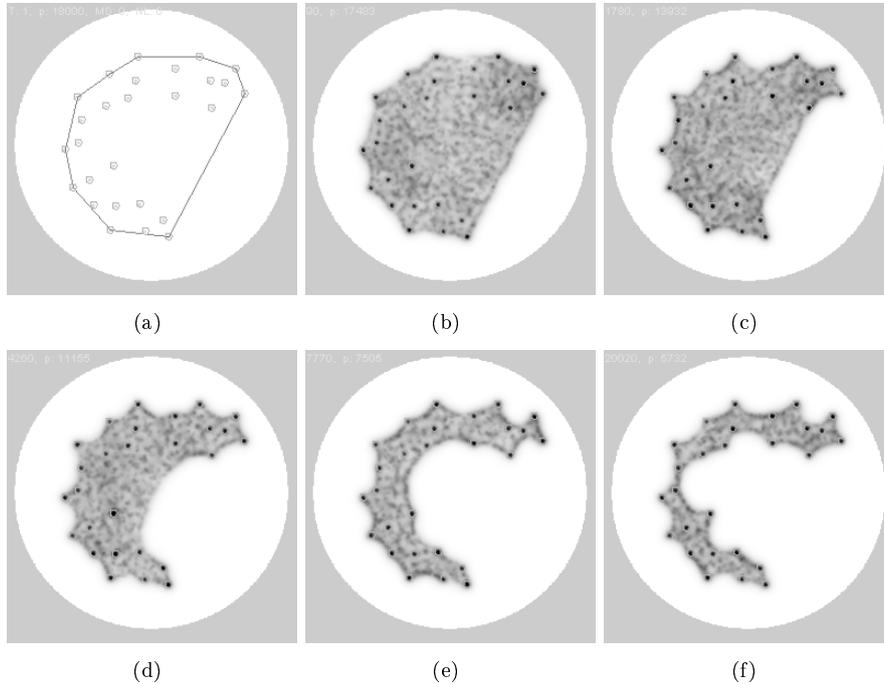

Fig. 25. Concave Hull by uniform shrinkage of the model plasmodium. (a) Set of points approximating the shape of letter 'C' cannot be intuitively represented by convex hull, (b-f) Approximation of concave hull by gradual shrinkage of the model plasmodium, $p$=18,000, $SA$ 60°, $RA$ 60°, $SO$ 7.

The graph of changing population size as the model plasmodium adapts (Fig. 26) shows that the population stabilises as the concave shape is adopted. If varying degrees of concavity are required, the current population size as a fraction of the original size, or alternatively the rate of population decline, could possibly be used as a simple parameter to tune the desired concavity.

The shrinkage of a solid mass of model plasmodium cannot construct $\alpha$-shapes, shapes with vacant regions within them, for example as with the letter '$A$'. However, by initialising the population at the node sites themselves, the individual fragments of 'plasmodium' fuse together and recover the general shape of the letter (Fig. 27,a-d). Further increasing the population size (manually or by biasing growth/shrinkage parameters) results in removal of the internal space and transition from an $\alpha$-shape to a solid Concave Hull (Fig. 27,e-f).

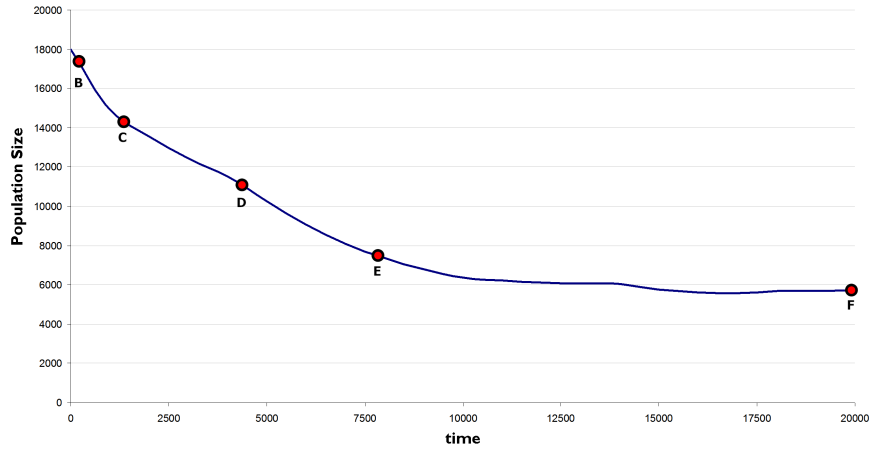

Fig. 26. Decrease in population size as concave shape formed. Population size over time. Letters and circles B-E represent corresponding images in Fig. 25(b–e).

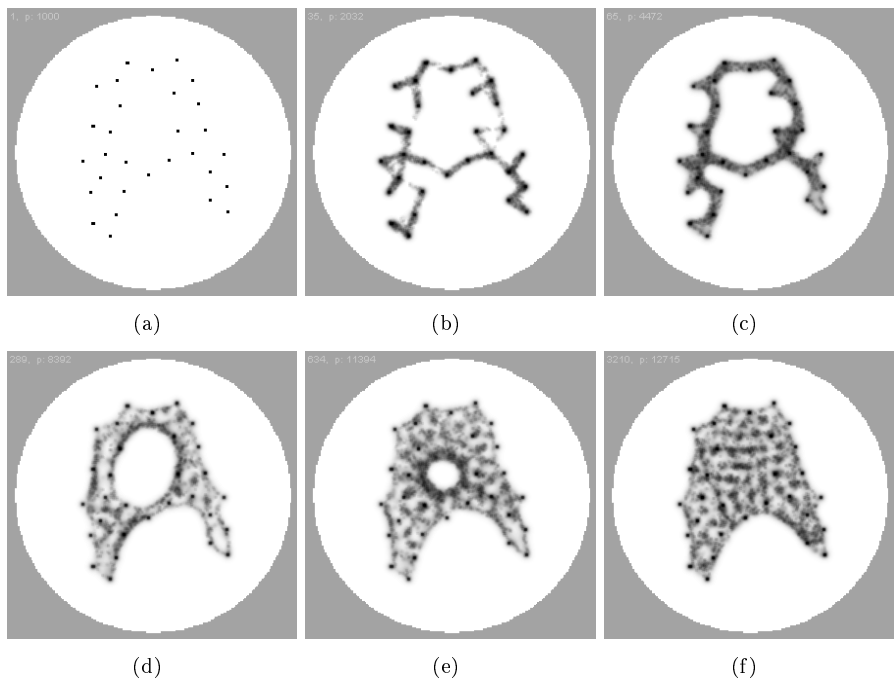

Fig. 27. Alternate method of generating $\alpha$-shape and Concave Hull by merging regions.

## 6.4. *Voronoi Diagrams and Variant Problems*

The Voronoi diagram of a set of $n$ points in the plane is the subdivision of the plane into $n$ cells so that every location within each cell is closest to the point within that cell. Conversely the bisectors forming the diagram are equidistant from the points between them. Computation of the Voronoi diagram may be achieved with a number of classical algorithms [35, 36] and are also the prototypical application solved by chemical reaction-diffusion non-classical computing devices [37, 38]. Non-classical approaches are based upon the intuitive notion of uniform propagation speed within a medium, emanating from the source nodes. The bisectors of the diagram are formed where the propagating fronts meet.

The method of using *P. polycephalum* to approximate Voronoi diagrams by avoidance of chemorepellents was described in [5, 26]. In this method a fully grown large plasmodium was first formed in a circular arena. Then repellent sources were introduced onto the plasmodium. The circular border of the arena was surrounded by attractants to maintain connectivity of the plasmodium network. The plasmodium then adapted its transport network to avoid the repellents whilst remain connected to the outer attractants, approximating the Voronoi diagram. Computation of Voronoi diagram may also be achieved by non-repellent methods. This method was proposed in [39] where plasmodia of *P. polycephalum* were inoculated at node sites on a nutrient-rich agar substrate. Attracted by the surrounding stimuli the plasmodia grew outwards in a radial pattern but when two or more plasmodia met they did not immediately fuse. There was a period where the growth was inhibited (presumably via some component of the plasmodium membrane or slime capsule) and the substrate at these positions was not occupied, approximating the Voronoi diagram. The position of the growth fronts remained stable before complete fusion eventually occurred.

The multi-agent model of *P. polycephalum* was used to approximate Voronoi diagrams in [24] and replicated both the repellent method (28), and the attractant methods (29). By varying the repellent concentration hybrid diagrams were formed which possessed features of both plane division and minimal object wrapping Fig. 30. Weighted Voronoi diagrams were approximated by varying the repellent size and concentration at different node sites (Fig. 31b). Hybrid Voronoi diagram and Proximity graph constructs were formed by using simultaneous placement of attractant and repellent sources 31c).

## 7. Summary and Conclusions

In this article we have examined the problem of how to influence, or program, spatially represented parallel unconventional computing substrates, inspired by research into the remarkable computational behaviour of true slime mould *Physarum polycephalum*. We used a multi-agent model of *P. polycephalum* which collectively behaves as a morphologically adaptive virtual material. We examined and catalogued low-level mechanisms which caused shape adaptation in the material. Some of these mechanisms were intrinsic to the parameter space of the model, in the form of pattern type and pattern scale. Other mechanisms (again, inspired by the effect of external stimuli on *P. polycephalum*) exerted their effect by changes in the environment. These included the addition and removal of nutrients, nutrient distance, angle and concentration, the presence of repellents, simulated light irradiation, and passive obstacles. These mechanisms may be used individually, or in combination to induce a wide range of morphological adaptation and approximate a range of spatially represented computing problems. Variants in the initial patterning, placement, and subsequent changing of problem stimuli may be used to cater for individual problem types, as can the level of external feedback control.

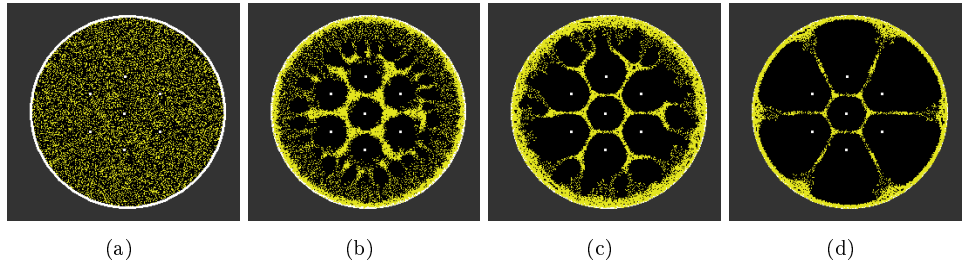

(a)                 (b)                 (c)                 (d)

Fig. 28. Approximation of Voronoi diagram by model in response to repulsive field. (a) Initial distribution of particles (yellow) representing a uniform mass of plasmodium, (b-c) particles respond to repulsive field by moving away from repellents, (d) final network connects outer attractant and bisectors correspond to Voronoi diagram.

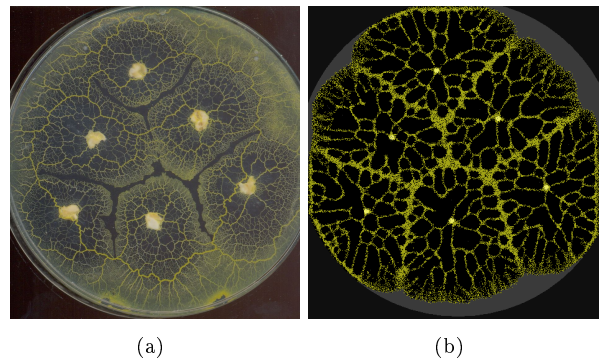

(a)                         (b)

Fig. 29. Approximation of Voronoi diagram by merging method. (a) Approximation of Voronoi diagram by *Physarum polycephalum*. Plasmodia are inoculated on oat flakes onto oatmeal agar. Expansive growth of plasmodia is inhibited at regions occupied by other plasmodia. These regions indicate bisectors of Voronoi diagram (Image courtesy of Andrew Adamatzky), (b) Attempt to reproduce bisector formation by model. The model plasmodium (particle positions shown) is not inhibited at regions of fusion, but bisector position is indicated by the increase in network density at these regions.

The approximation of a wide range of classical computing problems using parallel unconventional computing methods requires a set of computational tools, in much the same way that classical methods require different symbolic mechanisms (for example conditional program flow, execution branching, iterative mechanisms). Because unconventional computation is rooted in the physical world, these mecha-nisms typically affect the propagation of information throughout space. Components of the data sets (edges, vertices) may also be represented in the same space and be affected or manipulated by the spatial propagation, or indeed — as in the case of *P. polycephalum* — may manipulate the propagation of information themselves. Using tools inspired from the repertoire of slime mould we have presented mechanisms, which when combined, can approximate in parallel a wide range of computational tasks.

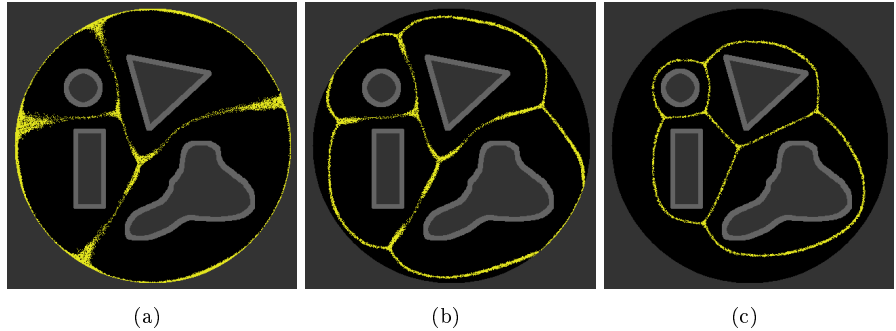

(a)          (b)          (c)

Fig. 30. Reducing repellent concentration allows minimising behaviour to exert its influence, inducing formation of hybrid Voronoi diagram. (a) at high concentration the repellent gradient forces the contractile network (yellow) to conform to the position of curved Voronoi bisectors between planar shapes, (b and c) reduction in repellent concentration allows contractile effects of transport network, minimising the connectivity between cells.

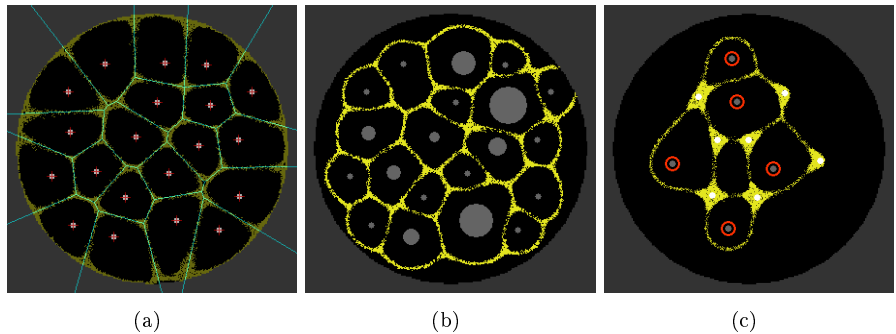

(a)          (b)          (c)

Fig. 31. Approximation of classical, weighted and hybrid Voronoi diagrams. a) classical Voronoi diagram (overlaid) is approximated by the model with repellent sources, b) weighted Voronoi diagram approximated using varying sized repellent stimuli, c) hybrid Voronoi and Proximity graph is approximated when both attractants (thick regions) and repellents (circled) are simultaneously presented to the model.

These tools may be used in an *ad hoc* manner to solve particular problem tasks, or they may provide the basis of a more formal grammar in order to delineate the range of problems which can be tackled by unconventional computing methods [9, 40]. The simplicity of the low-level mechanisms may also be attractive to distributed robotics applications in which robotic devices composed of simple components must be amenable to indirect guidance and external control whilst integrating a large number of distributed and concurrent inputs. In future work we will examine how these unconventional computing mechanisms can be combined and used to induce graph transformations in identical datasets.

## Acknowledgements

This work was supported by the EU research project "Physarum Chip: Growing Computers from Slime Mould" (FP7 ICT Ref 316366)

## References

[1] S. Takagi and T. Ueda. Emergence and transitions of dynamic patterns of thickness oscillation of the plasmodium of the true slime mold Physarum polycephalum. *Physica D*, 237:420–427, 2008.

[2] A. Takamatsu. Spontaneous switching among multiple spatio-temporal patterns in three-oscillator systems constructed with oscillatory cells of true slime mold. *Physica D: Nonlinear Phenomena*, 223(2):180–188, 2006.

[3] T. Nakagaki, H. Yamada, and A. Toth. Intelligence: Maze-solving by an amoeboid organism. *Nature*, 407:470, 2000.

[4] T. Nakagaki, R. Kobayashi, Y. Nishiura, and T. Ueda. Obtaining multiple separate food sources: behavioural intelligence in the *Physarum* plasmodium. *R. Soc. Proc.: Biol. Sci.*, 271(1554):2305–2310, 2004.

[5] T. Shirakawa, A. Adamatzky, Y.-P. Gunji, and Y. Miyake. On simultaneous construction of voronoi diagram and delaunay triangulation by *Physarum polycephalum*. *International Journal of Bifurcation and Chaos*, 19(9):3109–3117, 2009.

[6] A. Adamatzky. Developing proximity graphs by *Physarum polycephalum*: does the plasmodium follow the toussaint hierarchy? *Parallel Processing Letters*, 19:105–127, 2008.

[7] M. Aono and M. Hara. Amoeba-based nonequilibrium neurocomputer utilizing fluctuations and instability. In *6th Int. Conf., UC 2007*, volume 4618 of *LNCS*, pages 41–54, Kingston, Canada, August 13-17 2007. Springer.

[8] S. Tsuda, M. Aono, and Y.-P. Gunji. Robust and emergent *Physarum* logical-computing. *BioSystems*, 73:45–55, 2004.

[9] A. Adamatzky. *Physarum Machine*: Implementation of a Kolmogorov-Uspensky machine on a biological substrate. *Parallel Processing Letters*, 17(4):455–467, 2007.

[10] A. Adamatzky and J. Jones. Programmable reconfiguration of *Physarum* machines. *Natural Computing*, 9(1):219–237, 2010.

[11] S. Tsuda, K.-P. Zauner, and Y.-P. Gunji. Robot control with biological cells. *BioSystems*, 87:215–223, 2007.

[12] A. Adamatzky and J. Jones. Towards *Physarum* robots: computing and manipulating on water surface. *Journal of Bionic Engineering*, 5(4):348–357, 2008.

[13] A. Ishiguro, M. Shimizu, and T. Kawakatsu. A modular robot that exhibits amoebic locomotion. *Robotics and Autonomous Systems*, 54(8):641–650, 2006.

[14] J. Jones and A. Adamatzky. Emergence of self-organized amoeboid movement in a multi-agent approximation of *Physarum polycephalum*. *Bioinspiration and Biomimetics*, 7(1):016009, 2012.

[15] Y.-P. Gunji, T. Shirakawa, T. Niizato, M. Yamachiyo, and I. Tani. An adaptive and robust biological network based on the vacant-particle transportation model. *Journal of Theoretical Biology*, 272(1):187–200, 2011.

[16] A. Takamatsu, E. Takaba, and G. Takizawa. Environment-dependent morphology in plasmodium of true slime mold *Physarum polycephalum* and a network growth model. *Journal of theoretical Biology*, 256(1):29–44, 2009.

[17] A. Tero, R. Kobayashi, and T. Nakagaki. *Physarum* solver: A biologically inspired method of road-network navigation. *Physica A: Statistical Mechanics and its Applications*, 363(1):115–119, 2006.

[18] Y.-P. Gunji, T. Shirakawa, T. Niizato, and T. Haruna. Minimal model of a cell connecting amoebic motion and adaptive transport networks. *Journal of Theoretical Biology*, 253(4):659–667, 2008.

[19] D.S. Hickey and L.A. Noriega. Relationship between structure and information processing in *Physarum polycephalum*. *International Journal of Modelling, Identification and Control*, 4(4):348–356, 2008.

[20] D.R. Reyes, M.M. Ghanem, G.M. Whitesides, and A. Manz. Glow discharge in microfluidic chips for visible analog computing. *Lab Chip*, 2(2):113–116, 2002.

[21] S. Stepney. Embodiment. *In Silico Immunology*, pages 265–288, 2007.

[22] J. Jones. The emergence and dynamical evolution of complex transport networks from simple low-level behaviours. *International Journal of Unconventional Computing*, 6(2):125–144, 2010.

[23] J. Jones. Influences on the formation and evolution of *Physarum polycephalum* inspired emergent transport networks. *Natural Computing*, pages 1–25, 2011.

[24] J. Jones and A. Adamatzky. Slime mould inspired generalised voronoi diagrams with repulsive fields. *Int. J. Bifurcation and Chaos*, In-Press, 2013.

[25] J. Jones. Characteristics of pattern formation and evolution in approximations of *Physarum* transport networks. *Artificial Life*, 16(2):127–153, 2010.

[26] T. Shirakawa and Y.-P. Gunji. Computation of Voronoi diagram and collision-free path using the Plasmodium of *Physarum polycephalum*. *Int. J. Unconventional Computing*, 6(2):79–88, 2010.

[27] T. Latty and M. Beekman. Food quality affects search strategy in the acellular slime mould, *Physarum polycephalum*. *Behavioral Ecology*, 20(6):1160, 2009.

[28] A. Adamatzky. Physarum machines: encapsulating reaction–diffusion to compute spanning tree. *Naturwissenschaften*, 94(12):975–980, 2007.

[29] G.T. Toussaint. The relative neighbourhood graph of a finite planar set. *Pattern Recognition*, 12(4):261–268, 1980.

[30] J.A. Sellares and G. Toussaint. On the role of kinesthetic thinking in computational geometry. *International Journal of Mathematical Education in Science and Technology*, 34(2):219–237, 2003.

[31] R.A. Jarvis. On the identification of the convex hull of a finite set of points in the plane. *Information Processing Letters*, 2(1):18–21, 1973.

[32] A. Galton and M. Duckham. What is the region occupied by a set of points? *Geographic Information Science*, pages 81–98, 2006.

[33] M. Duckham, L. Kulik, M. Worboys, and A. Galton. Efficient generation of simple polygons for characterizing the shape of a set of points in the plane. *Pattern Recognition*, 41(10):3224–3236, 2008.

[34] H. Edelsbrunner, D. Kirkpatrick, and R. Seidel. On the shape of a set of points in the plane. *Information Theory, IEEE Transactions on*, 29(4):551–559, 1983.

[35] S. Fortune. A sweepline algorithm for voronoi diagrams. *Algorithmica*, 2(1):153–174, 1987.

[36] M. De Berg, O. Cheong, and M. Van Kreveld. *Computational geometry: algorithms and applications*. Springer-Verlag New York Inc, 2008.

[37] D. Tolmachiev and A. Adamatzky. Chemical processor for computation of voronoi

diagram. *Advanced Materials for Optics and Electronics*, 6(4):191–196, 1996.

[38] B. de Lacy Costello, N. Ratcliffe, A. Adamatzky, A.L. Zanin, A.W. Liehr, and H.G. Purwins. The formation of voronoi diagrams in chemical and physical systems: experimental findings and theoretical models. *International journal of bifurcation and chaos in applied sciences and engineering*, 14(7):2187–2210, 2004.

[39] A. Adamatzky. Physarum *Machines: Computers from Slime Mould*, volume 74. World Scientific Pub Co Inc, 2010.

[40] A. Schumann, K. Pancerz, and J. Jones. Towards logic circuits based on physarum polycephalum machines: The ladder diagram approach. In *Cliquet A., Plantier, G., Schultz, T., Fred, A., Gamboa H. (Eds.), Proceedings of the International Conference on Biomedical Electronics and Devices (BIODEVICES'2014)*, pages 165–170, Angers, France, March 3-6 2014. Springer.